\documentclass[aps,pre,twocolumn,preprintnumbers,showpacs,superscriptaddress,nobalancelastpage]{revtex4}

\begin{document}

\title{Network Synchronization in a Noisy Environment with Time Delays: Fundamental Limits and Trade-Offs}

\author{D. Hunt}
\affiliation{Department of Physics, Applied Physics, and Astronomy}

\author{G. Korniss\footnote{Corresponding author. korniss@rpi.edu}}
\affiliation{Department of Physics, Applied Physics, and Astronomy}

\author{B.K. Szymanski}
\affiliation{Department of Computer Science \\
Rensselaer Polytechnic Institute, 110 8th Street, Troy, NY 12180--3590, USA}

\begin{abstract}
We study the effects of nonzero time delays in stochastic
synchronization problems with linear couplings in an arbitrary
network. Using the known exact threshold value from the theory of
differential equations with delays, we provide the synchronizability
threshold for an arbitrary network. Further, by constructing the
scaling theory of the underlying fluctuations, we establish the
absolute limit of synchronization efficiency in a noisy environment
with uniform time delays, i.e., the minimum attainable value of the
width of the synchronization landscape.
Our results also have strong implications for optimization and
trade-offs in network synchronization with delays.
\end{abstract}


\date{\today}
\maketitle

\noindent
In network synchronization problems \cite{Arenas_PhysRep2008},
individual units, represented by nodes in the network, attempt to
adjust their local state variables (e.g., pace, load, orientation)
in a decentralized fashion. They interact or communicate only with
their local neighbors in the network, often with the intention to
improve global performance. These interactions or couplings can be
represented by directed or undirected, weighted or unweighted links.
Applications of the corresponding models range from physics,
biology, computer science to control theory, including
synchronization problems in distributed computing
\cite{GK_Science2003}, consensus, coordination and control in communication
networks
\cite{GK_PRE2007,Johari_IEEE2001,Saber_IEEE2004,Saber_IEEE2007},
flocking animals \cite{Vicsek_PRL1995,Cucker_IEEE2007}, bursting
neurons
\cite{Atay_PRL2003,Chen_EPL2008,Chen_PRE2009},
and cooperative control of vehicle formation \cite{Fax_IEEE2004}.

There has been a massive amount of research focusing on the
efficiency and optimization of synchronization problems
\cite{Arenas_PhysRep2008,Barahona_PRL2002,Nishikawa_PRL2002,LaRocca_PRE2008,LaRocca_PRE2009}
in various complex network topologies, including weighted
\cite{Zhou_PRL2006,GK_PRE2007} and directed
\cite{Saber_IEEE2007,Nishikawa_PRE2006,Nishikawa_2009}
networks. In this Letter, we study an aspect of stochastic
synchronization problems which is present in all real communication,
information, and computing networks
\cite{Saber_IEEE2004,Saber_IEEE2007,Huberman_IEEE1991,Strogatz_PRE2003},
including neurobiological networks \cite{Chen_EPL2008,Chen_PRE2009}:
the impact of {\em time delays} on synchronizability and on the
breakdown of synchronization.
The presence of time delays, however, will also present possible
scenarios for trade-offs. Here we show that when synchronization
networks are stressed by large delays, reducing local coordination
effort will actually improve global coordination. Similarly subtle
results have also been found in neurobiological networks with the
synchronization efficiency exhibiting non-monotonic behavior as a
function of the delay \cite{Chen_EPL2008,Chen_PRE2009}.

For our study, we consider the simplest stochastic model with linear
local relaxation, where network-connected agents locally adjust
their state to closely match that of their neighbors (e.g., load, or
task allocation) in an attempt to improve global performance.
However, they react to the information or signal received from their
neighbors with some time lag (as a result of finite processing,
queueing, or transmission delays), motivating our study of the
coupled stochastic equations of motion with delay,
\begin{equation}
\partial_{t} h_i(t) = - \sum_{j=1}^{N} C_{ij}[h_i(t-\tau_{ij})-h_j(t-\tau_{ij})] + \eta_{i}(t) \;.
\label{EW_synch_network}
\end{equation}
Here, $h_{i}(t)$ is the generalized local state variable on node $i$
and $\eta_{i}(t)$ is a delta-correlated noise with zero mean and
variance
$\langle\eta_{i}(t)\eta_{j}(t')\rangle$$=$$2D\delta_{ij}\delta(t-t')$,
where $D$ is the noise intensity.
$C_{ij}$$=$$C_{ji}$$\geq$$0$ is the symmetric coupling strength
($C_{ij}$$=$$W_{ij}A_{ij}$ in general weighted networks,
where $A_{ij}$ is the adjacency matrix and $W_{ij}$ is the link weight).
$\tau_{ij}$$>$$0$ is the time delay between two connected nodes $i$
and $j$. For initial conditions we use $h_i(t)$$=$$0$ for
$t$$\leq$$0$. Eq.~(\ref{EW_synch_network}) is also referred to as
the Edwards-Wilkinson process \cite{EW} on networks
\cite{GK_PRE2007} with time-delays. Without the noise term, the
above equation is often referred to as the consensus problem
\cite{Saber_IEEE2004,Saber_IEEE2007} on the respective network.

The standard observable in stochastic synchronization problems,
where relaxation competes with noise, is the width of the
synchronization landscape \cite{GK_Science2003,GK_PRE2007,LaRocca_PRE2008,LaRocca_PRE2009}
\begin{equation}
\langle w^2(t) \rangle \equiv
\left\langle\frac{1}{N}\sum_{i=1}^{N}[h_i(t)-\bar{h}(t)]^2\right\rangle\;,
\label{w2_def}
\end{equation}
where $\bar{h}(t)=(1/N)\sum_{i=1}^{N}h_i(t)$ is the global average of the local state variables and
$\langle\ldots\rangle$ denotes an ensemble average over the noise. A
network of $N$ nodes is synchronizable if $\langle w^2(\infty)
\rangle$$<$$\infty$, i.e., if the width approaches a finite value in the $t$$\to$$\infty$ limit.
The smaller the width, the better the synchronization.

In the case of uniform delays $\tau_{ij}\equiv\tau$, the focus of
this Letter, one can rewrite Eq.~(\ref{EW_synch_network}) as
\begin{equation}
\partial_{t} h_i(t) = - \sum_{j=1}^{N} \Gamma_{ij} h_j(t-\tau) + \eta_{i}(t)\;,
\label{EW_ntwk_gamma}
\end{equation}
where $\Gamma_{ij}=\delta_{ij}\sum_{l}C_{il}-C_{ij}$, is the
symmetric network Laplacian. In this case, by diagonalizing the
network Laplacian, one can decompose the problem into $N$
{\em independent} modes
\begin{equation}
\partial_{t} \tilde{h}_k(t) = - \lambda_k \tilde{h}_k(t-\tau) + \tilde{\eta}_{k}(t)\;,
\label{EW_synch_diag}
\end{equation}
where $\lambda_k$, $k=0,1,2,\ldots,N-1$, are the eigenvalues of the
network Laplacian and
$\langle\tilde{\eta}_{k}(t)\tilde{\eta}_{l}(t')\rangle$$=$$2D\delta_{kl}\delta(t-t')$.
For a single-component (or connected) network, the Laplacian has a
single zero mode (indexed by $k$$=$$0$) with $\lambda_0$$=$$0$, while
$\lambda_k$$>$$0$ for $k$$\geq$$1$.
Using the above eigenmode decomposition, the width of the synchronization landscape can be expressed
as $\langle w^2(t)\rangle = (1/N)\sum_{k=1}^{N-1}\langle \tilde{h}_k^2(t)\rangle$ \cite{GK_PRE2007}.

For example, for {\em zero time delay} ($\tau$$=$$0$), one immediately finds
$\langle w^2(t) \rangle$$=$$(1/N)\sum_{k=1}^{N-1} D\lambda_k^{-1}(1-e^{-2\lambda_k t})$.
The above expression explicitly shows that every finite connected
network with zero time delay is synchronizable, $\langle w^2(\infty) \rangle$$<$$\infty$.
In the limit of infinite network size, however, network ensembles
with a vanishing (Laplacian) spectral gap may become
unsynchronizable, depending on the details of the small-$\lambda$
behavior of the density of eigenvalues \cite{Arenas_PhysRep2008,GK_Science2003,GK_PRE2007}.

\begin{figure}[t]
\vspace*{-1.20truecm}
\centering
\includegraphics{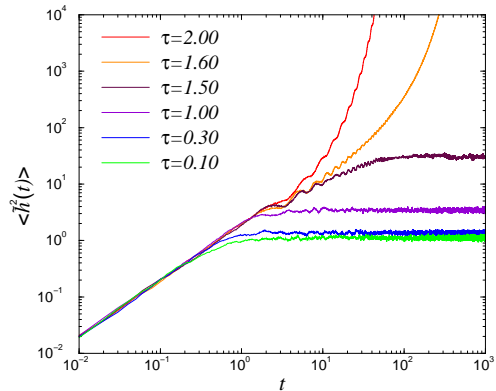}
\vspace*{6.30truecm}
\caption{
Time series $\langle\tilde{h}^2(t)\rangle$ for different delays, obtained by
numerically integrating Eq.~(\ref{noise_delay_eq}) and averaging over $1,000$ independent realization of the noise.
Here, $\lambda$$=$$1$, $D$$=$$1$, and $\Delta t$$=$$0.01$.
The theoretical (continuum-time) threshold value of the delay
(for $\langle\tilde{h}(\infty)\rangle$ to remain bounded)
is $\tau_{\rm c}$$=$$\pi/2$ \cite{note_numerical}.}
\label{h2_time}
\vspace*{-0.40truecm}
\end{figure}

In the case of {\em non-zero uniform delays}, the case considered here,
the eigenmodes of the problem are again governed by a
stochastic equation of motion of identical form for all $k$$\geq$$1$
[Eq.~(\ref{EW_synch_diag})]. Thus, understanding the time-evolution
of a single stochastic variable $\tilde{h}_k(t)$ and its
fluctuations will provide both full insight to the synchronizability
condition of the network-coupled system and a framework to compute
the width of the synchronization landscape. Therefore, to ease notational burden and
to direct our focus to a single stochastic variable, we will temporarily drop the
index $k$ referring to a specific eigenmode, and study the
stochastic differential equation
\begin{equation}
\partial_{t} \tilde{h}(t) = - \lambda \tilde{h}(t-\tau) + \tilde{\eta}(t)
\label{noise_delay_eq}
\end{equation}
with $\langle\tilde{\eta}(t)\tilde{\eta}(t')\rangle$$=$$2D\delta(t-t')$.
Using standard Laplace transformation with initial conditions
$\tilde{h}(t)$$=$$0$ for $t$$\leq$$0$, one finds
\begin{equation}
\tilde{h}(t) =
\int_{0}^{t} dt'\tilde{\eta}(t') \sum_{\alpha}\frac{e^{s_\alpha(t-t')}}{1+\tau s_\alpha} \;,
\label{delay_diff_eq_solution}
\end{equation}
where $s_\alpha$, $\alpha=1,2,\ldots$, are the solutions of the characteristic equation
\begin{equation}
s + \lambda e^{-\tau s} = 0 \;
\label{complex_eq}
\end{equation}
in the complex plane. The above complex equation has an {\em
infinite} number of solutions for $\tau$$>$$0$
\cite{Frisch_1935,Hayes_1950,Saber_IEEE2004}.
Using Eq.~(\ref{delay_diff_eq_solution}), for the noise-averaged
fluctuations we find
\begin{equation}
\langle \tilde{h}^2(t)\rangle =
\sum_{\alpha,\beta} \frac{-2D(1-e^{(s_\alpha + s_\beta) t})}{(1+\tau s_\alpha)(1+\tau s_\beta)(s_\alpha + s_\beta)}  \;.
\label{h2_tau}
\end{equation}
The solution of Eq.~(\ref{complex_eq}) with the largest real part governs the long-time
temporal behavior of the respective mode (e.g., stability, approach
to, or relaxation in the steady state). The condition for
$\langle\tilde{h}^2(\infty)\rangle$ to remain finite is ${\rm
Re}(s_\alpha)$$<$$0$ for all $\alpha$. As has been shown for
Eq.~(\ref{complex_eq}), this inequality holds if
$\tau\lambda$$<$$\pi/2$
\cite{Frisch_1935,Hayes_1950,Saber_IEEE2004}.
In Fig.~\ref{h2_time} we show the time-dependent width of the fluctuations associated
with a single stochastic variable,
obtained by numerically integrating Eq.~(\ref{noise_delay_eq}) for
a few characteristic cases \cite{note_numerical}.

Returning to the context of network synchronization,
synchronizability requires a finite steady-state width, $\langle
w^2(\infty)\rangle = (1/N)\sum_{k=1}^{N-1}\langle
\tilde{h}_k^2(\infty)\rangle<\infty$. Thus, for uniform time delays
in a given network, all $k$$\geq$$1$ modes must have finite
steady-state fluctuations $\langle
\tilde{h}_k^2(\infty)\rangle<\infty$. This implies that one must
have $\tau\lambda_k$$<$$\pi/2$ for all $k$$\geq$$1$ modes, or
equivalently \cite{consensus_threshold},
\begin{equation}
\tau\lambda_{\rm max}<\pi/2 \;.
\label{synch_condition}
\end{equation}
The above exact delay threshold for synchronizability has some
immediate and profound consequences for unweighted networks. Here,
the coupling matrix is identical to the adjacency matrix,
$C_{ij}$$=$$A_{ij}$, and the bounds and the scaling properties of
the extreme eigenvalues of the network Laplacian are well known. In
particular, \mbox{$Nk_{\rm max}/(N-1)$$\leq$$\lambda_{\rm
max}$$\leq$$2k_{\rm max}$} \cite{Anderson_1985}, where
$k_{\rm max}$ is the maximum node degree in the network (i.e.,
$\langle\lambda_{\rm max}\rangle$$=$${\cal O}(\langle k_{\rm
max}\rangle)$). Thus, $\tau k_{\rm max}$$<$$\pi/4$ is sufficient for
synchronizibility \cite{consensus_threshold}, while $\tau k_{\rm
max}$$>$$\pi/2$ leads to the breakdown of synchronization with
certainty.  These inequalities imply that networks with potentially
large degrees, e.g., scale-free (SF) networks
\cite{Barab_sci,BarabREV}, are rather vulnerable to intrinsic
network delays \cite{Saber_IEEE2004,Saber_IEEE2007}.
For example, SF network ensembles with a natural degree cut-off
exhibit $\langle\lambda_{\rm max}\rangle$$\sim$$\langle k_{\rm
max}\rangle$$\sim$$N^{1/(\gamma-1)}$ for $N$$\gg$$1$ (when the
average degree $\langle k\rangle$ is held fixed), where $\gamma$ is
the exponent governing the power-law tail of the degree distribution
\cite{MendesREV}. In turn, the probability that a
realization of a random SF network ensemble of $N$ nodes is
synchronizable approaches zero for {\em any nonzero} delay $\tau$
in the limit of $\tau N^{1/(\gamma-1)}$$\gg$$1$.

\begin{figure}[t]
\vspace*{-1.20truecm}
\centering
\includegraphics{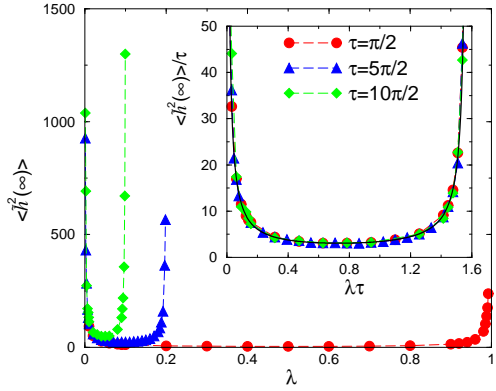}
\vspace*{6.30truecm}
\caption{Steady-state fluctuations $\langle\tilde{h}(\infty)\rangle$ obtained by
numerically integrating Eq.~(\ref{noise_delay_eq}) as a function of $\lambda$ for different $\tau$ values.
Here, $D$$=$$1$, and $\Delta t$$=$$0.01$. The inset shows the scaled plot of the same data points,
$\langle \tilde{h}^2(\infty)\rangle/\tau$ vs $\lambda\tau$, together with the
numerically fitted scaling function $f(x)$ (solid curve) \cite{note_numerical}.}
\label{fig_h2_lambda_tau}
\vspace*{-0.40truecm}
\end{figure}

Next, we analyze the steady-state behavior of the width in the
synchronizable regime. We accomplish this by investigating the basic
scaling features of the steady-state fluctuations of a single
stochastic variable, $\langle \tilde{h}^2(\infty)\rangle$, governed
by Eq.~(\ref{noise_delay_eq}), which can be associated with an
arbitrary mode. In this regime one must have ${\rm
Re}(s_\alpha)$$<$$0$ for all $\alpha$, or equivalently, $\tau\lambda$$<$$\pi/2$.
Defining a new variable $z$$\equiv$$\tau s$, Eq.~(\ref{complex_eq}) can
be rewritten $z + \lambda\tau e^{-z}=0$, i.e., for a given $\lambda$
and $\tau$, the solutions for the scaled variable $z$ can only
depend on $\lambda\tau$, $z_\alpha$$=$$z_\alpha(\lambda\tau)$,
$\alpha=1,2,\ldots$. Thus, the solutions of the characteristic equation
Eq.~(\ref{complex_eq}) must exhibit the scaling form
$s_\alpha=\tau^{-1} z_\alpha(\lambda\tau)$, $\alpha=1,2,\ldots$.
Substituting the above expression into Eq.~(\ref{h2_tau}) and taking
the $t$$\to$$\infty$ limit immediately yields the scaling form
\begin{equation}
\langle \tilde{h}^2(\infty)\rangle = D\tau f(\lambda\tau) \;.
\label{scaling_fx}
\end{equation}
Thus, for a single stochastic variable $\tilde{h}(t)$ governed by
the stochastic differential equation Eq.~(\ref{noise_delay_eq})
(simple relaxation in a noisy environment with delay), plotting
$\langle \tilde{h}^2(\infty)\rangle/\tau$ vs $\lambda\tau$ (for
a fixed noise intensity $D$) should yield full data collapse, as demonstrated in Fig.~\ref{fig_h2_lambda_tau} \cite{note_numerical}.
[While we do not have an analytic expression for the scaling
function, for small arguments it asymptotically has to scale as
$f(x)$$\simeq$$1/x$ to reproduce the exact limiting case of zero
delay, $\langle \tilde{h}^2(\infty)\rangle$$\simeq$$D/\lambda$.
Further, we numerically found that in the vicinity of $\pi/2$, it
approximately diverges as $(\pi/2-x)^{-1}$.]
The scaling function $f(x)$ is clearly non-monotonic; it exhibits a
single minimum, at approximately $x^*$$\approx$$0.73$ with
$f^*$$=$$f(x^*)$$\approx$$3.1$. The immediate message of the above
result is rather interesting:
For a single stochastic variable governed by Eq.~(\ref{noise_delay_eq})
with a nonzero delay, there is an optimal value of the relaxation coefficient
$\lambda^*$$=$$x^*/\tau$, at which point the steady-state
fluctuations attain their minimum value $\langle
\tilde{h}^2(\infty)\rangle$$=$$D\tau f^*$$\approx$$3.1 D\tau$. This
is in stark contrast with the zero-delay case where $\langle
\tilde{h}^2(\infty)\rangle$$=$$D/\lambda$, i.e., there the steady-state
fluctuation is a monotonically decreasing function of the relaxation
coefficient.

In addition to gaining fundamental insights, constructing the
scaling function $f(x)$ numerically with some acceptable precision
of the single variable problem (Fig.~\ref{fig_h2_lambda_tau} inset)
also provides a method to obtain the steady-state width of the network-coupled system: one can
numerically diagonalize the Laplacian of the underlying network and
employ the scaling function $f(x)$ to obtain the width,
\begin{equation}
\langle w^2(\infty) \rangle =
\frac{1}{N}\sum_{k=1}^{N-1}\langle \tilde{h}_k^2(\infty)\rangle =
\frac{D\tau}{N}\sum_{k=1}^{N-1} f(\lambda_k\tau)\;.
\label{w2_method}
\end{equation}
Further, we can now extract the minimum attainable width of the
synchronization landscape in a noisy environment with uniform time
delays. For a fixed $\tau$, each term in Eq.~(\ref{w2_method}) can
be minimized by choosing $\lambda_{k}=x^*/\tau$ for all
$k$$\geq$$1$. Then
\begin{equation}
\langle w^2(\infty) \rangle^* = \frac{N-1}{N} D \tau f^* \approx 3.1 D\tau \;
\label{w2_limit}
\end{equation}
for large $N$. This number, the fundamental limit of synchronization
efficiency in a noisy environment with uniform time delays, can be
used as a base-line value when comparing networks from the viewpoint
of synchronization efficiency. Note that there is a trivial network
which realizes the optimal behavior: the fully connected graph with
identical coupling constants $C_{ij}$$=$$x^*/N\tau$ for all
$i$$\neq$$j$. (This network has $N$$-$$1$ identical non-zero
eigenvalues, $\lambda_{k}$$=$$x^*/\tau$ for all $k$$\geq$$1$.) In
general, networks with a narrow spectrum centered about
$\lambda^*=x^*/\tau$ shall perform closer to optimal. How to
construct such networks with possible topological and cost
constraints is a different and challenging question which we will
not pursue in detail here, but we note that essentially the same
problem arises in the broader context of synchronization of
generalized dynamical systems \cite{Nishikawa_PRE2006,Nishikawa_2009}. Recent methods
tackling this issue involve locally reweighting and/or removing links
from the networks to achieve optimal performance
\cite{Nishikawa_2009}.

The essential non-monotonic feature of the scaling function $f(x)$
in Eq.~(\ref{w2_method}) (including the potentially
diverging contributions from large eigenvalues beyond the threshold)
presents various trade-off scenarios in network synchronization
problems with delays. As the simplest and obvious application of the
above results, consider a network which is stressed by large delays
beyond its threshold, $\tau\lambda_{\rm max}$$>$$\pi/2$ (so that the largest
fluctuations and the width are growing exponentially without bound).
Then even a suitably chosen uniform reduction of all couplings
$C_{ij}'$$=$$pC_{ij}$ ($\lambda_k'$$=$$p\lambda_k$)
with $p$$<$$\pi/2\lambda_{\rm max}\tau$ will lead to the
stabilization of the system, with a finite steady-state width.
\begin{figure}[t]
\vspace*{-1.20truecm}
\centering
\includegraphics{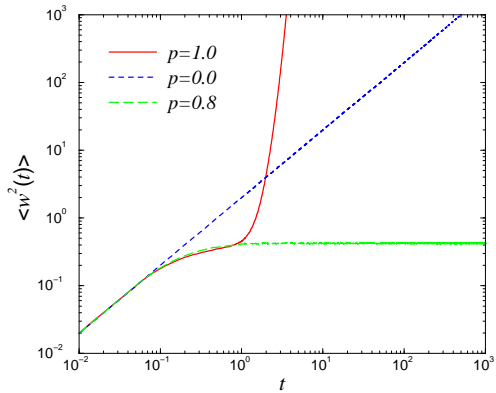}
\vspace*{6.30truecm}
\caption{
Time dependent width for $\tau$$=$$1.2\pi/2\lambda_{\rm max}$ for different values of the local synchronization rate $p$ on a fixed graph,
obtained by the numerical integration of Eq.~(\ref{EW_synch_network}) with $\Delta t$$=$$0.005$ and $D$$=$$1.0$.
The underlying network is a Barab\'asi-Albert SF graph \cite{Barab_sci}
with $N$$=$$100$, $\langle k\rangle$$\approx$$6$, and $\lambda_{\rm max}$$\approx$$32$.}
\label{w2_prob}
\vspace*{-0.40truecm}
\end{figure}
In communication and computing networks, the effective coupling
strength $C_{ij}$ can be controlled by the frequency (or rate) of local
synchronizations through the respective link \cite{GK_Science2003}. The above results then
suggest that when the system is beyond its stability threshold,
synchronizing sufficiently {\em less frequently}, can lead to
stabilization and better coordination.
Figure~\ref{w2_prob} shows results for the case when the communication neighborhood is fixed,
but the local synchronizations through the links [the coupling terms in
Eq.~(\ref{EW_synch_network})] are only performed with
probability $p$$\leq$$1$, while invoking the noise term at every time step.
Indeed, reducing the local synchronization rate can improve global performance.
In fact, even performing no local synchronizations at all ($p$$=$$0$) leads to a slower power-law
divergence of the width with time, $\langle w^2(t)\rangle $$\simeq$$2Dt$,
as opposed to the exponential divergence governed by the largest
eigenvalue(s) above the threshold.

In summary, we have obtained the delay threshold for the simplest
stochastic synchronization problem with linear couplings in an
arbitrary network. Further, by exploring and investigating the
scaling properties of the fluctuations associated with the
eigenmodes of the network Laplacian, we found the minimum attainable
steady-state width of the synchronization landscape in any network.
The non-monotonic feature of the scaling function governing the
fluctuations can guide potential trade-offs and optimization in
network synchronization.
For systems with more general (non-linear) node dynamics, one can
also expect that the synchronizability phase diagram will exhibit
non-monotonic behavior as a function of the coupling strength and/or
the delays
\cite{Chen_EPL2008,Chen_PRE2009,Huberman_IEEE1991,Strogatz_PRE2003}.
In real communication and information networks, the delays
$\tau_{ij}$ are not uniform \cite{Johari_IEEE2001},
but are affected by the network neighborhood and spatial distance.
We currently investigate the impact of heterogeneous delays on
network synchronization.

We thank S. Sreenivasan and T. Caraco for discussions and a careful
reading of the manuscript. This work was supported in part by DTRA
Award No. HDTRA1-09-1-0049 and by the Army Research Laboratory under
Cooperative Agreement Number W911NF-09-2-0053. The views and
conclusions contained in this document are those of the authors and
should not be interpreted as representing the official policies,
either expressed or implied, of the Army Research Laboratory or the
U.S. Government.



\begin{thebibliography}{100}
\bibitem{note_numerical}
The time discretization of Eq.~(\ref{noise_delay_eq}), and in
general, that of Eq.~(\ref{EW_ntwk_gamma}) naturally has its own
effects on the stability through the numerical scheme. Choosing
$\Delta$$t$$<<$$\tau$ and  $\Delta$$t$$<<$$1/\lambda_{\rm max}$ will
yield only small corrections to the behavior of the underlying
continuous-time system.
\end{thebibliography}
\end{document}